\documentclass[12pt,preprint]{aastex}

\usepackage{natbib}

\begin{document}

\title{An Evolving Stellar Initial Mass Function and the Gamma-Ray Burst Redshift Distribution}

\author{F. Y. Wang\altaffilmark{1,2} and  Z. G. Dai\altaffilmark{1,2}}

\affil{$^1$Department of Astronomy, Nanjing University, Nanjing
210093, China \\
$^2$Key laboratory of Modern Astronomy and Astrophysics (Nanjing
University), Ministry of Education, Nanjing 210093, China}

\begin{abstract}
Recent studies suggest that \emph{Swift} gamma-ray bursts (GRBs) may
not trace an ordinary star formation history. Here we show that
the GRB rate turns out to be consistent with the star formation
history with an evolving stellar initial mass function (IMF). We
first show that the latest \emph{Swift} sample of GRBs reveals an
increasing evolution in the GRB rate relative to the ordinary star
formation rate at high redshifts. We then assume only massive stars
with masses greater than the critical value to produce GRBs, and use
an evolving stellar IMF suggested by Dav\'{e} (2010) to fit the
latest GRB redshift distribution. This evolving IMF would increase
the relative number of massive stars, which could lead to more GRB
explosions at high redshifts. We find that the evolving IMF can well
reproduce the observed redshift distribution of \emph{Swift} GRBs.

\end{abstract}


\section{Introduction}
Gamma-ray bursts (GRBs) are brief flashes of $\gamma$-rays occurring
at an average detection rate of a few events per day at cosmological
distances. Because of their very high luminosity, GRBs can be
detected out to the edge of the visible Universe (Ciardi \& Loeb
2000; Lamb \& Reichart 2000; Bromm \& Loeb 2002; Gou et al. 2004).
Thus, GRBs are ideal tools for probing the star formation rate, the
reionization history, and the metal enrichment history of the
Universe (Totani 1997; Campana et al. 2007; Bromm \& Loeb 2007). The
advantages of GRBs over quasars for probing the high-redshift
Universe had been discussed by Bromm \& Loeb (2007). In addition,
GRBs have been used as standard candles to constrain cosmological
parameters and dark energy (Dai, Liang \& Xu 2004; Friedman \& Bloom
2005; Wang \& Dai 2006; Schaefer 2007, and references therein).

The association of long GRBs with core-collapse supernovae naturally
suggests that the cosmic GRB rate should trace the star formation
history. This gave rise to the expectation that GRBs may be a good
tracer of cosmic star formation (Totani 1997; Wijers et al. 1998;
Lamb \& Reichart 2000; Blain \& Natarajan 2000; Porciani \& Madau
2001). However, it was found that the rate of GRBs increases with
cosmic redshift faster than the ordinary star formation rate (SFR)
does (Daigne et al. 2006; Le \& Dermer 2007; Kistler et al. 2008,
2009; Y\"{u}ksel \& Kistler 2007; Cen \& Fang 2007; Li 2008; Wang \&
Dai 2009; Butler et al. 2010; Wanderman \& Piran 2010). The reason
for this discrepancy has been unknown.

By investigating the redshift distribution of \emph{Swift} GRBs,
Guetta \& Piran (2007) found that the observed high-redshift bursts
are more than the expectation from an ordinary star formation
history (SFH) and thus the high-redshift GRB rate is inconsistent
with the one inferred from the current model for the SFR.
Furthermore, Kistler et al. (2008) found that the GRB rate at
redshift $z\simeq 4$ is about four times larger than expected from
star formation measurements. Daigne et al. (2006) concluded that GRB
properties or progenitors must evolve with cosmic redshift to
reconcile the observed GRB redshift distribution with the measured
SFH. Li (2008) explained the observed discrepancy between the GRB
rate history and the star formation rate history as being due to
cosmic metallicity evolution, by assuming that long GRBs tend to
occur in galaxies with low metallicities. However, very recently
Levesque et al. (2010a,b) found several high-metallicity long GRB
host environments, which suggests that a low-metallicity cut-off is
unlikely (also see Graham et al. 2009). Xu \& Wei (2008) used a
factitious stellar initial mass function (IMF) evolving with
redshift to interpret the GRB redshift distribution. Cheng et al.
(2010) suggested that this discrepancy could be eliminated if some
high-redshift GRBs are ascribed to electromagnetic bursts of
superconducting cosmic strings, although the existence of the
superconducting cosmic strings has remained controversial.

In this Letter, we first enlarge the GRB sample with 122 long GRBs
observed by \emph{Swift}. Then we interpret the latest \emph{Swift}
GRB redshift distribution using a reasonable evolving stellar
initial mass function (IMF) proposed by Dav\'{e} (2010). The
structure of this paper is as follows: in section 2, we give an
evolving initial mass function form, and in section 3, we show the
analysis method. The results are presented in section 4 and
conclusions are shown in section 5.

\section{An evolving initial mass function}
The ordinary form of stellar IMF proposed by Salpeter (1955) is
\begin{equation}
dN= m^{-1.35}d\log m.
\end{equation}
It was shown that the high-redshift GRB rate exceeds the expectation
based on the above SFR. This leads us to invoking a different form of
stellar IMF. The possibility of an evolving IMF was discussed
several times in the literature (e.g. Larson 1998, 2005; Ferguson,
Dickinson \& Papovich 2002; Fardal et al. 2007; van Dokkum 2008).
Kroupa (2001) pointed out that a universal IMF is not expected
theoretically, though no variations had been unequivocally detected
in the studies of local star-forming regions. Scalo (1998) mentioned
that although the IMF index may vary at different redshifts in the
Universe, its average value is close to the Salpeter value. Wilkins,
Trentham \& Hopkins (2008) independently determined the cosmic
stellar mass growth rate by compiling observations of stellar mass
densities from the literature and suggested an evolving IMF to
interpret the discrepancy between the stellar mass density and SFH.
In order to reconcile the discrepancy between the theory predicting
the galaxy stellar mass-SFR relation with the observations, Dav\'{e}
(2008) proposed an evolving IMF with the following form
\begin{equation}
\frac{dN}{d\log{m}}=\xi(m)\propto\left\{
\begin{array}{l}
m^{-0.3}\;\; {\rm for}\; m<\hat{m}_{\rm IMF}\\
m^{-1.3}\;\; {\rm for}\; m>\hat{m}_{\rm IMF},
\end{array} \right.
\end{equation}
where $\hat{m}_{\rm IMF}=0.5 (1+z)^{2} M_\odot$, which has been
constrained by requiring non-evolving star formation activity
parameter. It is worth noting that this evolving IMF is only
constrained out to $z\sim 2$ from the galaxy stellar mass-SFR
relation, though its predictions are consistent with the other
observations out to $z\sim 4$. Chary (2008) found that this IMF
would produce sufficient ionizing photons to account for late
reionization of the intergalactic medium if it evolved out to $z>4$.
So the IMF suggested by Dav\'{e} (2008) can be used to $z>4$. More
recently, Dav\'{e} (2010) found $\hat{m}_{\rm IMF}=0.5
(1+z)^{3-0.75z} M_\odot$ using the Herschel data.

\section{The method}
We consider a spatial volume $V$ at redshift $z$. The IMF can be
written as $A\xi(m)$, so $\int_{m_s }^{m_l } {Am\xi (m)d{\rm{log }}m
= R_{\rm SFR} V}$, where $R_{\rm SFR}$ is the SFR, $m_l$ is the
largest mass of stars and $m_s$ is the smallest mass of stars. We
consider only massive stars with masses larger than $30M_{\odot}$
can produce GRBs (Woosley 1993; Bissaldi et al. 2007)\footnote{The
lower limit mass of a star that can collapse to GRB is uncertain at
present. But this value is unimportant in our analysis below. The
best fitting parameters will shift slightly when the
lower limit mass is changed. But this evolving IMF could still interpret the
GRB redshift distribution.}, so
\begin{equation}
R_{\rm GRB}  \propto \frac{{N_{m > 30M_ \odot  } }}{V} =K\left(
{\frac{c}{{H_0 }}} \right)^{ - 3} \frac{{\int_{30M_ \odot  }^{m_l }
{\xi (m)d\log m} }}{{\int_{m_s }^{m_l } {m\xi (m)d\log m} }}R_{\rm
SFR},
\end{equation}
where $K$ is a constant to be constrained and $R_{\rm GRB}$ is the
rate of GRBs, representing the number of GRBs per unit time per unit
volume at redshift $z$. We use the SFR derived by Hopkins \& Beacom
(2006),
\begin{eqnarray}
    \log R_{\rm SFR}(z) = a+b \log(1+z) \;,   \label{sfr}
\end{eqnarray} with \begin{eqnarray}
    (a,b) = \left\{\begin{array}{ll}
        (-1.70,3.30) \;, & z<0.993 \\
        (-0.727,0.0549) \;, & 0.993<z<3.80 \\
        (2.35,-4.46) \;, & z>3.80
        \end{array}\right. \;
    \label{ab}
\end{eqnarray}
Le Borgne et al. (2009) used mid- and far-infrared observations to
constrain the SFR and found that SFH is well-constrained and
consistent with direct measurements from Hopkins \& Beacom (2006).
So we also use the results of Hopkins \& Beacom (2006).

Then the observed rate of GRBs within $z\sim z+dz$ and $L\sim
L+dL$ is
\begin{equation}
\frac{{dN}}{{dt}} = \Phi (L)\frac{{R_{\rm GRB} }}{{1 +
z}}\frac{\Delta \Omega_s}{4\pi}\frac{{dV_{\rm com}(z)}}{{dz}}dLdz,
\end{equation}
where $\Phi (L)$ is the beaming-convolved luminosity function of
GRBs, $(1+z)^{-1}$ is due to cosmological time dilation and $\Delta \Omega_s=1.4$~sr
is the solid angle covered on the sky by \emph{Swift} (Salvaterra \& Chincarini 2007). In a flat
universe, the comoving volume is calculated by
\begin{eqnarray}
    \frac{dV_{\rm com}}{d z} = 4\pi D_{\rm com}^2 \frac{dD_{\rm com}}{d z} \;,
\end{eqnarray}
where the comoving distance is
\begin{eqnarray}
    D_{\rm com}(z) \equiv \frac{c}{H_0} \int_0^z \frac{dz^\prime}{\sqrt{
            \Omega_m(1+z^\prime)^3 + \Omega_\Lambda}} \;.
\end{eqnarray}
In the calculations, we use $\Omega_m=0.3$, $\Omega_\Lambda=0.7$ and
$H_0$=70 km~s$^{-1}$~Mpc$^{-1}$. There are many luminosity function
forms in the literature. We use the Schechter-function form
\begin{equation}
\Phi (L) = \frac{1}{{L_ \star  }}\left( {\frac{L}{{L_ \star  }}}
\right)^\beta  \exp ( - L/L_ \star  ),
\end{equation}
where $\beta$ and $L_\star$ are constant parameters to be determined
by the observational data. The observed distribution of $L_{\rm
iso}$ is then given by
\begin{equation}
\Delta N(L) = \Phi (L)\left[ {\int_0^{z_{\max } (L)} {\frac{{R_{\rm
GRB} (z)}}{{1 + z}}\frac{\Delta \Omega_s}{4\pi}\frac{{dV_{\rm com}}}{{dz}}dz} } \right]\Delta
L\Delta t_{\rm obs},
\end{equation}
where $z_{\rm max} = z_{\rm max}(L_{\rm iso})$ is the maximum
redshift up to which a GRB with luminosity $L_{\rm iso}$ can be
detected by \emph{Swift}, solved from equation $L_{\rm lim} (z) =
L_{\rm iso}$.

The isotropic-equivalent luminosity of a GRB can be obtained by
$L_{\rm iso}=E_{\rm iso}(1+z)/T_{90}$ (Kistler et al. 2008). We use
122 long GRBs observed by \emph{Swift}\footnote{See $\rm
http://swift.gsfc.nasa.gov/docs/swift/archive/grb\_table.$} till GRB
090726 (Butler et al. 2010). The distribution of $L_{\rm iso}$ for
the 122 GRBs in the sample is shown in Fig.1. The luminosity
threshold can be approximated by a bolometric energy flux limit
$F_{\lim} = 1.2 \times 10^{-8}$erg cm$^{-2}$ s$^{-1}$. The
luminosity threshold is then
\begin{eqnarray}
    L_{\lim} = 4\pi D_L^2 F_{\rm lim} \;,   \label{lum_lim}
\end{eqnarray}
where $D_L$ is the luminosity distance to the burst.

With the above luminosity threshold and an adopted GRB rate history,
the observed luminosity distribution can be fitted by an intrinsic
Schechter luminosity function with a power-law index $\beta=-1.12$,
a characteristic luminosity $L_ \star=9.16 \times 10^{52}$ erg
s$^{-1}$ and $\Delta t_{\rm obs}K=69858.51$ with $\chi^2_r=1.15$.

\section{The results}
In order to study the rates of GRBs and star formation, it is
convenient to use a dimensionless $Q$, where $Q=Q(z)$ is defined by
(Kistler et al. 2007)
\begin{eqnarray}
    Q(z) \equiv \left(\frac{c}{H_0}\right)^{-3} \int_0^z
        \frac{1}{1+z^\prime}\frac{dV_{\rm com}}{d z^\prime} dz^\prime \;.
    \label{Q}
\end{eqnarray}
The coordinate $Q$ is particularly useful in binning the data, as
the definition of $Q$ has taken into account both the effect of the
comoving volume and the effect of cosmic time dilation. For example,
when the comoving rate density of GRBs was a constant, in each
equally sized bin of $Q$, the observed GRBs number would be a
constant. The complete GRB selection function is very difficult to
determine (Coward 2007). We choose the GRBs in the cuts $L_{\rm
iso}>0.8\times10^{51}$erg s$^{-1}$ and in the redshift range $0-4$
(Kistler et al. 2008). This method can reduce the selection effect
by removing many low-$z$, low-$L_{\rm iso}$ bursts that could not
have been seen at higher redshifts. There are 72 GRBs in this
sample. The SFR fit from Hopkins \& Beacom (2006) in this range is
shown as the dotted line in Fig. 2. We can see the GRB rate is
incompatible with the expectation from the ordinary SFR. The solid
line shows the cumulative distribution of the 72 \emph{Swift} GRBs.
The result from an evolving IMF suggested by Dav\'{e} (2010) is
shown as the solid line, which agrees with the observed data very
well.

From Eqs (8), (11) and (14), we can obtain the observed number of
GRBs in an observer's time interval $\Delta t_{\rm obs}$, and with
$Q$ in the interval $Q-(Q + dQ)$,
\begin{equation}
\Delta N(Q) = \left[ {\int_{L_{\lim } (z)}^\infty  {\Phi
(L)\frac{{R_{\rm GRB} (z)}}{{1 +
z}}\frac{\Delta \Omega_s}{4\pi}\frac{{dV(z)}}{{dz}}\frac{{dz}}{{dQ}}dL} } \right]dQ\Delta t_{\rm
obs}, \label{NQ}
\end{equation}
where $L_{\rm lim}(z)$ is determined by Eq.~(\ref{lum_lim}).

We use the luminosity cut $L_{\rm lim}=0.8\times10^{51}$erg s$^{-1}$
(Li 2008). The cut in luminosity and redshift minimizes the
selection effect in the GRB data. The total number of GRBs with
$L_{\rm iso}>L_{\rm lim}$ is 72. The distribution of $Q$ for the 72
\emph{Swift} GRBs is plotted in Fig. 3. Because
of the flux limit of the detector (Kistler et al. 2008), the model
deviates from the data at $z>4$. The solid line shows the
best fit of the $R_{\rm GRB}$ to the first six data points using an evolving IMF.
The dashed line shows the best fit from non-evolving IMF.

Fig. 4 shows the distribution of $Q$ for all 122 \emph{Swift} GRBs
in the sample. The solid line is $N(Q)$ calculated by Eq.~\ref{NQ}
with the normalization and the luminosity function parameters
determined above, and $L_{\rm lim}$ calculated by Eq.~\ref{lum_lim}.
We can see that the modeled $N(Q)$ fits the observational data very
well with $\chi^2_r=1.14$. However, there is an obvious excess in
the number of GRBs in the bin of $0 < Q < 1$, which might be caused
by statistical fluctuations. If we exclude it, the $\chi^2_r$
decreases to 0.32.

\section{Conclusions}
In this Letter, we have presented that the redshift distribution of
\emph{Swift} GRBs with measured redshifts and calculated
luminosities can be successfully fitted by the SFH with an evolving
stellar IMF. It is widely considered by current theories that only
massive stars with masses larger than the critical value can produce
long GRBs. The evolving stellar IMF becoming increasingly top heavy
at larger $z$ suggested by Dav\'{e} (2010) can lead to more GRBs
produced at high redshifts.

Kistler et al. (2008) considered several possible reasons
for the discrepancy between the \emph{Swift} GRB rate and the SFH.
They showed that the Kolmogorov-Smirnov test does not favor an
interpretation as a statistical anomaly. Selection effects are also
unlikely to cause an increased efficiency in detecting high-redshift
GRBs. Although Kistler et al. (2008) have argued that alternative
reasons are possible (e.g., evolution in the fraction of binary
systems, an evolving IMF of stars, cosmic metallicity evolution),
they did not give a quantitative analysis or a detailed discussion
of the evolving IMF. We enlarged the GRB sample with 122 long GRBs
and used a reasonable evolving IMF. The results in this paper
indicate that the evolving IMF may explain the redshift distribution
of \emph{Swift} GRBs. If the redshift distribution of GRBs and SFH
are well measured, GRBs would be used to probe the stellar IMF at
high redshifts.

\acknowledgements We thank D. M. Wei for valuable discussions
and an anonymous referee for useful
suggestions and comments. This work is supported by the National Natural Science Foundation of
China (grants 10873009 and 11033002) and the National Basic Research
Program of China (973 program) No. 2007CB815404.

\begin{figure}
\plotone{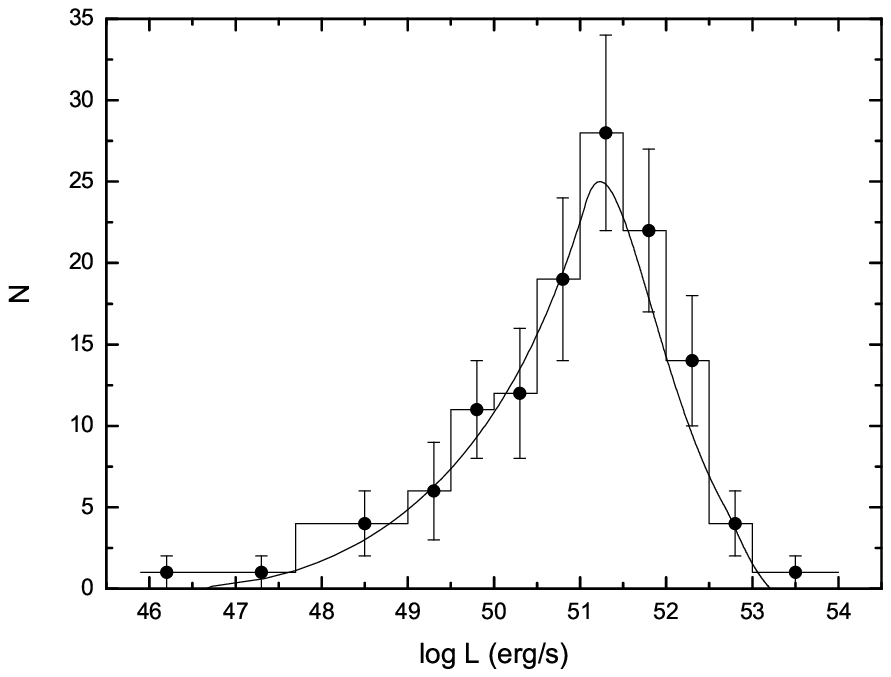} \caption{Distribution of the
isotropic-equivalent luminosity for 122 long-duration \emph{Swift}
GRBs. The solid line is plotted according to Eq. 12 using best
fitted parameters. }
\end{figure}

\begin{figure}
\plotone{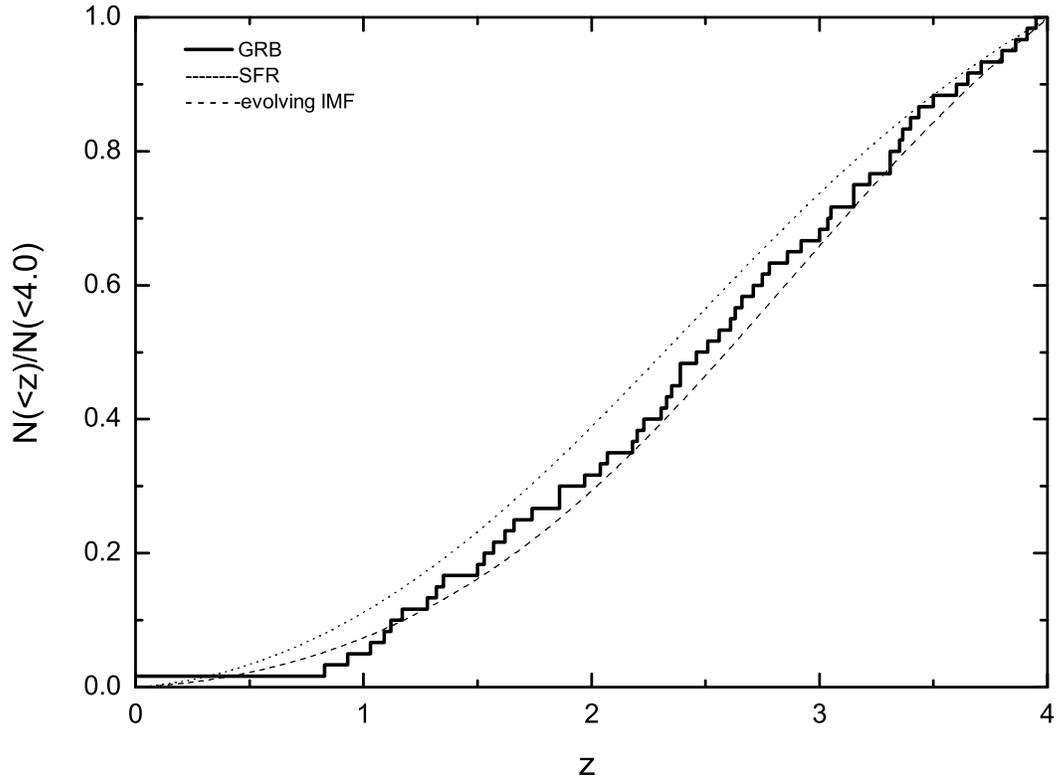} \caption{The cumulative distribution of
72 \emph{Swift} long GRBs with $L_{\rm iso}>0.8\times10^{51}$ erg
s$^{-1}$ (stepwise solid line). The dotted line shows the GRB rate
inferred from the star formation history of Hopkins \& Beacom
(2006). The dashed line shows the GRB rate inferred from star
formation history including an evolving IMF.}
\end{figure}

\begin{figure}
\plotone{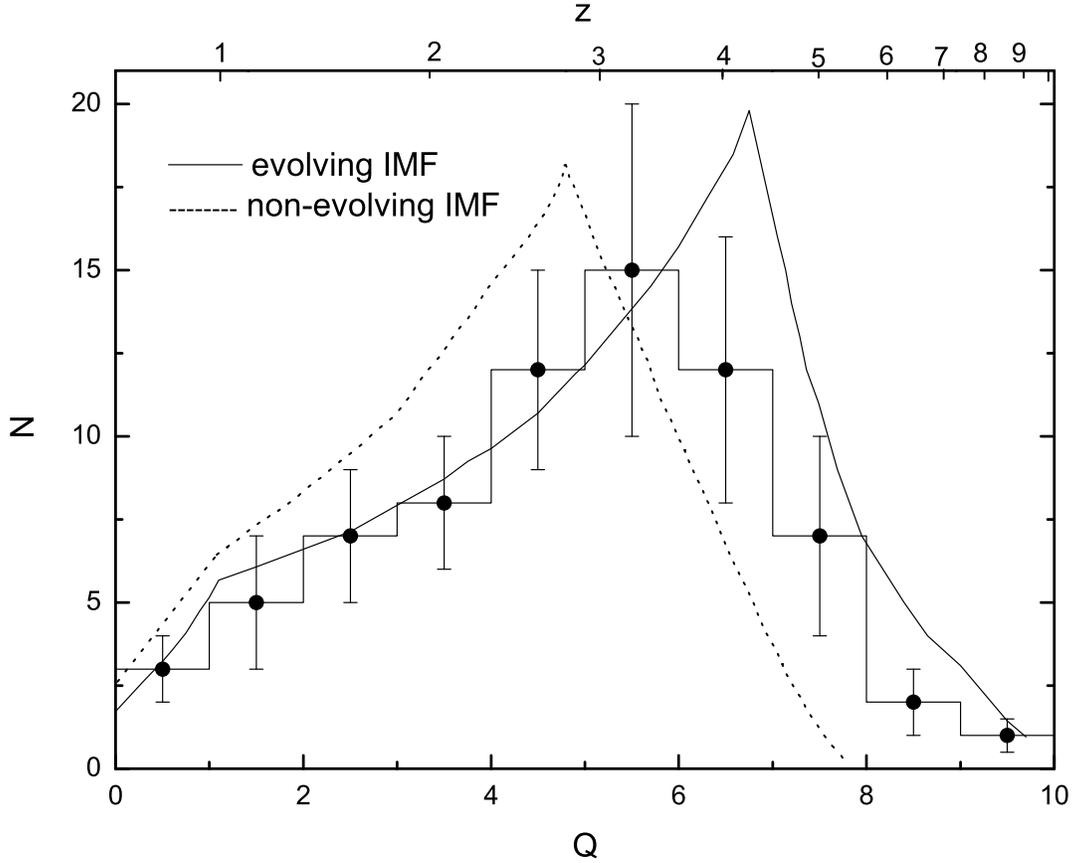} \caption{Distribution of $Q$ for 72
\emph{Swift} GRBs with $L_{\rm iso}>0.8\times 10^{51}$erg s$^{-1}$
(the solid histogram, with the number of GRBs in each bin indicated
by a dark point with Poisson error bars). The solid line is the best
fit of the GRB rate. The dotted curve shows the best fit by
non-evolving IMF.}
\end{figure}

\begin{figure}
\plotone{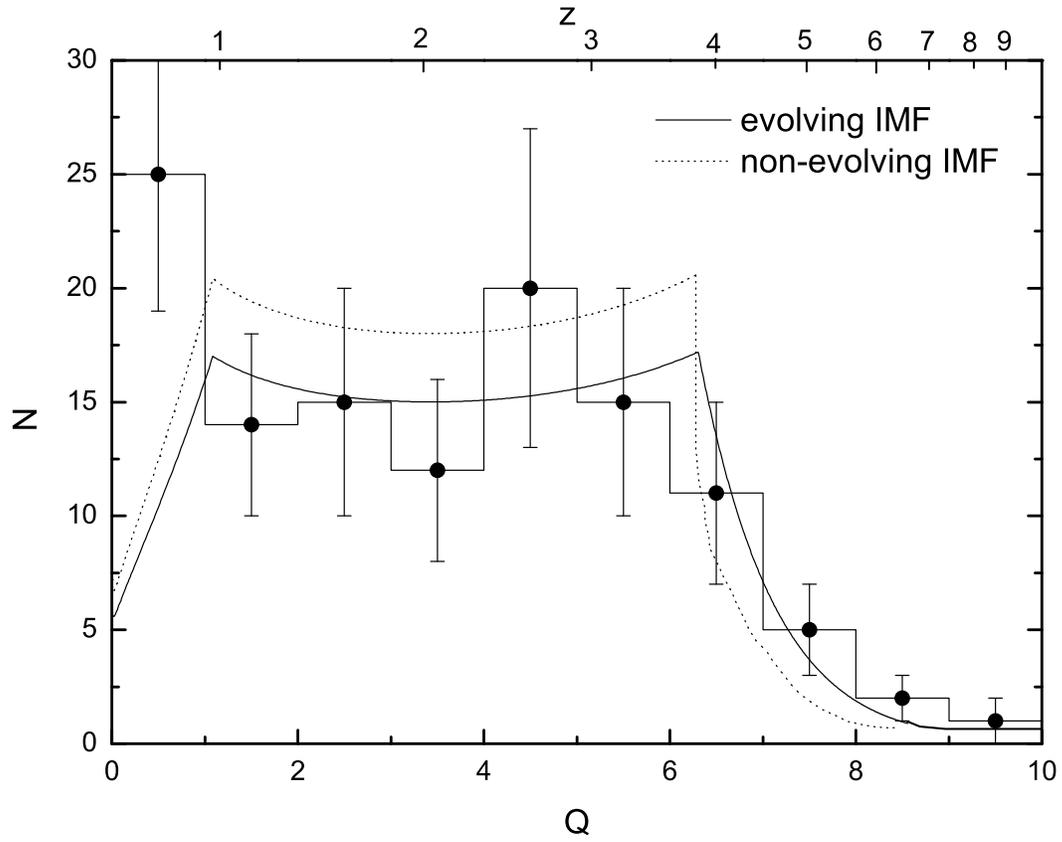} \caption{Distribution of $Q$ for all the 122
\emph{Swift} GRBs, the points with error bars represent the number
of GRBs lying between $Q_i \sim Q_i + 1$. The solid curve shows the
best fit by Eq. 13. The dotted curve shows the best fit by
non-evolving IMF.}
\end{figure}

\end{document}